\documentclass[sigconf]{acmart}

\AtBeginDocument{%
  \providecommand\BibTeX{{%
    \normalfont B\kern-0.5em{\scshape i\kern-0.25em b}\kern-0.8em\TeX}}}

\setcopyright{acmcopyright}

\copyrightyear{2022}
\acmYear{2022}
\setcopyright{othergov}\acmConference[AM '22]{AudioMostly 2022}{September 6--9, 2022}{St. Pölten, Austria}
\acmBooktitle{AudioMostly 2022 (AM '22), September 6--9, 2022, St. Pölten, Austria}
\acmPrice{15.00}
\acmDOI{10.1145/3561212.3561244}
\acmISBN{978-1-4503-9701-8/22/09}

\usepackage{times}
\usepackage{latexsym}

\usepackage[T1]{fontenc}

\usepackage[utf8]{inputenc}

\usepackage{microtype}

\usepackage{graphicx}
\usepackage[inline]{enumitem}
\usepackage{booktabs}
\usepackage{braket}
\usepackage{threeparttable}
\usepackage{multirow}
\usepackage{tikz}
\usepackage{afterpage}
\usepackage{makecell}
\usepackage{float}
\usepackage{stfloats}

\begin{document}

\title{What Did I Just Hear? Detecting Pornographic Sounds in Adult Videos Using Neural Networks}


  
\author{Holy Lovenia}
\authornote{The work was done when the author was working in Prosa.ai.}
\affiliation{%
  \institution{Prosa.ai}
  \country{}}
\email{holy.lovenia@gmail.com}

\author{Dessi Puji Lestari}
\affiliation{%
  \institution{Bandung Institute of Technology}
  \institution{Prosa.ai}
  \country{}
  }

\author{Rita Frieske}
\affiliation{%
  \institution{The Hong Kong University of Science and Technology (HKUST)}
  \country{}}


\renewcommand{\shortauthors}{Holy Lovenia, Dessi Puji Lestari, Rita Frieske}

\begin{abstract}
  Audio-based pornographic detection enables efficient adult content filtering without sacrificing performance by exploiting distinct spectral characteristics. To improve it, we explore pornographic sound modeling based on different neural architectures and acoustic features. We find that CNN trained on log mel spectrogram achieves the best performance on Pornography-800 dataset. Our experiment results also show that log mel spectrogram allows better representations for the models to recognize pornographic sounds. Finally, to classify whole audio waveforms rather than segments, we employ voting segment-to-audio technique that yields the best audio-level detection results.
\end{abstract}

\ccsdesc[500]{Computing methodologies~Artificial intelligence}

\keywords{Pornographic detection, sound modeling, neural networks}

\begin{teaserfigure}
  \centering
  \includegraphics[width=0.9\textwidth, trim={0, 3cm, 0, 3cm}, clip]{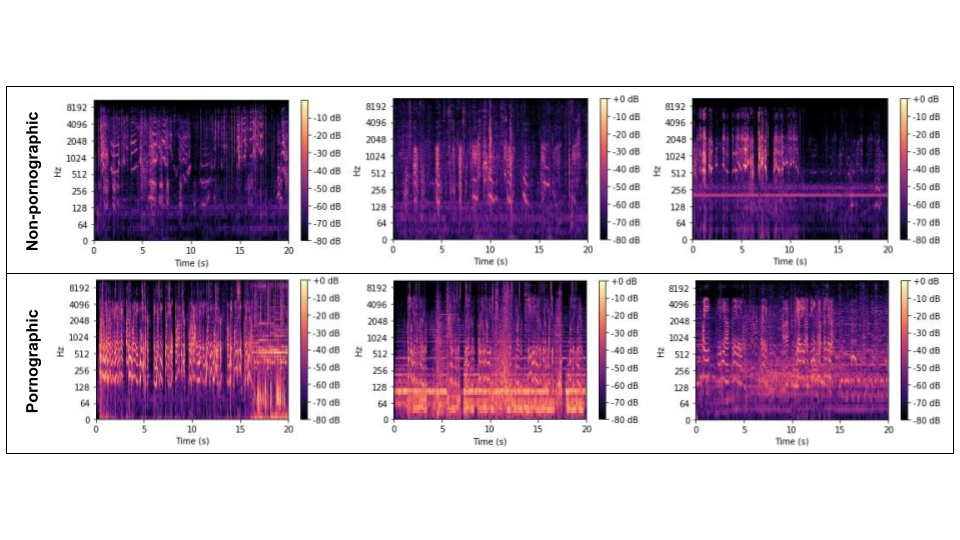}
  \caption{Log-frequency power spectrograms of \textbf{non-pornographic} and \textbf{pornographic} sound signals.}
  \Description{Log-frequency power spectrograms of non-pornographic and pornographic sound signals.}
  \label{fig:spec}
\end{teaserfigure}

\maketitle

\section{Introduction}
\label{intro}

Parental filters are a necessity to ensure safe internet use by young internet users~\cite{hornor2020child,dwulit2019prevalence,stanley2018pornography}. The sheer number of online materials causes controlled filtering and distribution of mature content to be impossible, and even internet activities with a non-sexual intention could lead to unintentional exposure to sexually explicit videos~\cite{hornor2020child}.

Employing machine learning, multiple studies have worked on automatic adult video detection, with the majority of these focusing on visual classification techniques~\cite{perez2017video,tang2009pornprobe,avila2013pooling}. In general, the performance of vision-based detection heavily relies on the quality, quantity, and diversity of the visual data. Images and videos with visual deterioration (e.g., poor lighting levels, obstruction, blurry objects, and out-of-focus scenes) are hard to classify~\cite{lim2011automatic, Karamizadeh2018}. In addition, several vision-based methods tend to regard images or videos containing a large amount of skin color as pornographic, because exposed skin is prevalent in sexually explicit content~\cite{kim2010automatic,Tabone2021}. Data-related hindrances aside, vision-based approaches also need a substantial memory capacity to save the model parameters and computation power to process large features~\cite{lim2011automatic}.

In comparison to vision-based approaches, research regarding adult video detection based on acoustic features is underdeveloped. Audio-based approaches benefit from requiring a relatively lower amount of space and computational resources in development due to their smaller number of features. Audio-based classifications can also exploit distinct spectral characteristics that discriminate the audios of adult videos from those of non-adult videos. Visual examples of the first 20 seconds of pornographic and non-pornographic audio data are provided in Figure \ref{fig:spec}.

In this work, we present successful methods of pornographic sound detection using neural architectures and compare the advantages of different acoustic features. Furthermore, we also analyse segment-to-audio prediction methods to classify whole audios based on their comprising segments.

\section{Related work}
\label{sec:related-works}

Different from speech tasks like automatic speech recognition~\cite{yu2022automatic} or speech synthesis~\cite{dhiaulhaq2021}, classifying audio in pornographic videos is treated as a specific sound recognition problem.
Log mel spectrograms and mel frequency cepstral coefficients (MFCCs) are the most widely used auditory features to solve diverse tasks of sound recognition, e.g., environmental and animal sounds \cite{su2019environment, chung2013automatic, mitrovic2006discrimination}.

A number of studies have specifically worked on pornographic sound detection using various methods: MFCCs and vector quantization~\cite{qu2011}; non-syllable-based 2D MFCCs (STDMFCCs) and a Gaussian mixture model (GMM)~\cite{kim2010automatic}; support vector machines (SVMs) and repeated curve-like spectrum features (RCSFs)~\cite{lim2011automatic}; the Radon transform and GMM~\cite{kim2011automatic}; MFCCs and pitch-based features~\cite{Banaeeyan2019}; MFCCs and long-term short memory (LSTM)~\cite{wazir2019acoustic}; and most recently log-mel spectrograms~\cite{app11073066}.
Most attempt to improve pornographic sound detection by learning the optimal feature representations, or employing conventional machine learning models. As a result, pornographic sound modeling with deep learning has not been properly explored despite the notable results recently achieved by deep neural networks (DNNs), such as feedforward neural networks (FFNNs) and convolutional neural networks (CNNs)~\cite{li2017comparison,hershey2017cnn}.

\section{Methodology}
\label{sec:method}

This section describes an overview of the three main aspects of our pipeline: the types of acoustic features used (\S\ref{ssec:acoustic-features}), the neural model architectures (\S\ref{ssec:nn-architectures}), and the methods we use to derive audio-level classification results from the corresponding segment-level predictions (\S\ref{ssec:segment-to-audio}). We enable both segment-level and audio-level detection because the former can be used to filter adult videos during streaming and the latter to filter adult videos offline.

\subsection{Acoustic features}
\label{ssec:acoustic-features}

Since audio data contain not only human speech, but also other noises, we need acoustic features that could discriminate pornographic from  non-pornographic audio data. We use MFCCs, which are widely used for speech detection, and log mel spectrograms, which contain more acoustic information and thus retain more artifacts in high and low frequencies. Such artifacts preserve sounds that are not typical of speech (e.g., moans or sounds of moving bodies), but are important for pornographic audio detection. Both MFCCs and log mel spectrograms are state-of-the-art features used for speech recognition, but their application to pornographic audio detection is still a novelty.

\subsection{Neural architectures}
\label{ssec:nn-architectures}

The DNN architecture has become the state-of-the-art method in many machine learning tasks, such as automatic speech recognition (ASR) and computer vision, and has replaced other methods in visual pornography detection~\cite{Tabone2021,Karamizadeh2018}. In this work, we employ an FFNN and a CNN. The use of the latter method is justified by its success in ASR. The convolutions in CNNs perform operations similar to parametric feature extraction. Hence, they are capable of identifying sounds that are outside the typical speech range, but might be crucial for pornographic sound detection \cite{Loweimi2019}.

\subsection{Segment-to-audio prediction methods}
\label{ssec:segment-to-audio}

We use four methods to derive a whole audio waveform's classification result from its corresponding segment-level predictions.

The first is a \textbf{standard threshold}, where an audio waveform's prediction probability equals the average of the respective segments' prediction probabilities. An audio waveform is pornographic (class 1) if the probability is greater than the threshold, or non-pornographic (class 0) otherwise. We use a fixed threshold of 0.5.

The second method is a \textbf{validation-dependent threshold}, which resembles the standard threshold method. However, instead of using a fixed threshold, each model establishes the threshold based on its audio-level performance on the validation set. A threshold of $\{0, 0.1, \ldots, 0.9\}$ is alternately used by the model to make predictions on the validation data. Afterwards, the threshold that yields the highest validation performance will be employed in the evaluation.

The third method is the \textbf{segment majority}, which labels an audio waveform as pornographic if the number of pornographic segments is higher than the number of non-pornographic segments.

Finally, the fourth method is \textbf{voting}. If an audio is denoted as pornographic by both the validation-dependent threshold and segment majority, then it is voted as pornographic.

\section{Pornographic audio dataset}
\label{sec:data-dev}



We acquire 800 videos (400 pornographic and 400 non-pornographic) from the Pornography-800 dataset.\footnote{https://sites.google.com/site/pornographydatabase/}
After excluding one video that has no audio from the dataset, we extract the audio waveforms from the remaining videos.
The dataset statistics are described in Table \ref{tab:data-description}.

\begin{table}[b]
    \centering
    \resizebox{0.7\columnwidth}{!}{%
    \begin{tabular}{*3r}
         & \textbf{Porn.} & \textbf{Non-porn.} \\
        \toprule
        \textbf{Duration (minutes)} & & \\
        Sum & 3,400 & 1,200 \\
        Mean & 8.5 & 3 \\
        Max & 33.5 & 20 \\
        Min & 1 & 1 \\
        \midrule
        \textbf{\# Data sample (files)} & 399 & 400 \\
        \bottomrule
    \end{tabular}
    }
    \caption{Statistical overview of Pornography-800 dataset. Durations are rounded for brevity.}
    \label{tab:data-description}
\end{table}


\paragraph{Audio characteristics}
\label{ssec:data-characteristics}

Upon manual inspection, we find that the pornographic audio data generally contain human sounds provoked by sexual activities, such as moans, deep breathing, panting, groans, screams, and whimpers. Other comprising elements are silences, various environmental sounds (e.g., bed creaking, sheets rustling, or movement noise), background music, and conversations. The non-pornographic audio data mainly consist of background music and human speech (not necessarily conversational). Since diverse activities are present, the non-pornographic audio data also comprise rustling noises, babies' cries, engine hums, audience cheers, and indiscernible sounds. Although the original dataset marks half of the non-pornographic data as “difficult” to classify for the vision-based model~\cite{avila2013pooling}, there is no significant difference between the sound compositions of the “easy” and “difficult” non-pornographic audio data. While many vision-based detection models could be misled into incorrect prediction by the amount of skin shown in the “difficult” non-pornographic videos, sound-based models might perform better in pornography detection due to the sound compositions in the “easy” and “difficult” non-pornographic videos being consistent.

\section{Pornographic sound experiment}
\label{sec:classification-exp}

This section focuses on experiments employing two feature extraction methods and two neural models.


\subsection{Preprocessing and feature extraction}
\label{ssec:data-prep}



We resample all audio data to 16 kHz and split it into training, validation, and test sets at the approximate ratio of 60:20:20. All splits are class-balanced.
Considering the wide audio duration range, we slice the audio data into overlapping segments with a hop length of 1 second. We employ 20-second and 60-second segment lengths, and each segment obtains its label from its original audio.

We extract MFCCs (12 coefficients) and log mel spectrograms (26 mel bands) from the 20-second and 60-second segment data, obtaining four feature sets: 1) mfcc-20, 2) mfcc-60, 3) log-mel-20, and 4) log-mel-60.
We apply cepstral mean and variance normalization (CMVN) in all data splits, utilizing the average and variance
of each coefficient from the training set \cite{molau2003feature}.

\subsection{Models, training settings, and evaluation}
\label{ssec:model-building}
\label{ssec:train-eval}


We employ an FFNN and CNN to experiment on pornographic sound classification. The FFNN's dense layers consist of 32, 16, and 1 units. Convolution layers in the CNN contain 32 filters with a kernel size of 2 and a stride of 1. We use rectified linear unit (ReLU) and sigmoid as activation functions in both models to introduce non-linearity to the neural networks. We build the models using the Keras\footnote{https://github.com/fchollet/keras} library.

We use a batch size of 128 and employ the Adam optimizer, which iteratively adjusts the network weights to minimize loss during the training process, with an adaptive learning rate of 1e-3 to 1e-4.
As our pornographic sound detection is a binary classification task, we use binary cross-entropy as the loss function.
We apply class weighting to alleviate the imbalanced data's effect on the models. We train the CNN and FFNN using each feature set derived from the segmented data (\S\ref{ssec:data-prep}). We evaluate the performances using the mean F1-score on the segment level and audio level. To derive audio-level predictions from the segment-level predictions, we employ the methods described in \S\ref{ssec:segment-to-audio}.

\section{Results and discussion}
\label{sec:results}

We describe the experimental results of each model and feature set pair on the validation and test data in \S\ref{ssec:classification-result}, and provide our analysis of the incorrect audio-level detection in \S\ref{ssec:incorrect-detection-analysis}.

\subsection{Pornographic sound detection results}
\label{ssec:classification-result}

\begin{table}[t]
    \centering
    \resizebox{0.85\columnwidth}{!}{%
    \begin{tabular}{c l *2r}
        \toprule
        \textbf{Model} & \textbf{Feature set} & \textbf{Valid F1-score} & \textbf{Test F1-score} \\
        \toprule
        \multirow{4}{*}{\vspace{-1mm} \bfseries{FFNN}} & mfcc-20 & 89.56\% & 90.25\% \\
        & mfcc-60 & 91.89\% & 90.99\% \\
        & log-mel-20 & 92.88\% & 92.49\% \\
        & log-mel-60 & 93.72\% & 93.42\% \\
        \midrule
        \multirow{4}{*}{\vspace{-1mm} \bfseries{CNN}} & mfcc-20 & 93.38\% & 92.46\% \\
        & mfcc-60 & 93.46\% & \underline{93.92\%} \\
        & log-mel-20 & \bfseries{95.73\%} & 93.70\% \\
        & log-mel-60 & \underline{95.38\%} & \bfseries{94.89\%} \\
        \bottomrule
    \end{tabular}
    }
    \caption{Segment-level performance comparison between acoustic features, MFCCs and log mel spectrograms, for pornographic sound detection.}
    \label{tab:result-per-segment}
\end{table}

\begin{table*}[t]
    \centering
    \resizebox{1.55\columnwidth}{!}{%
    \begin{tabular}{c l *4c}
        \toprule
        \multirow{2}{*}{\vspace{-2mm}\bfseries{Model}} & \multirow{2}{*}{\vspace{-2mm}\bfseries{\makecell{Feature set}}} & \multicolumn{4}{c}{\bfseries{Test F1-score (per audio)}} \\
        \cmidrule{3-6}
         & & \bfseries{\makecell{Standard threshold}} & \bfseries{\makecell{Segment majority}} & \bfseries{\makecell{Val-dependent threshold}} & \bfseries{Voting} \\
        \toprule
        \multirow{4}{*}{\bfseries{FFNN}} & mfcc-20 & 83.04\% & 83.53\% & 84.75\% & 83.53\% \\
        & mfcc-60 & 87.27\% & 87.27\% & 85.71\% & 87.27\% \\
        & log-mel-20 & 88.10\% & 87.43\% & 88.20\% & 88.20\% \\
        & log-mel-60 & 89.44\% & 89.44\% & 89.44\% & 89.44\% \\
        \midrule
        \multirow{4}{*}{\bfseries{CNN}} & mfcc-20 & 86.71\% & 86.71\% & 87.72\% & 87.72\% \\
        & mfcc-60 & 87.21\% & 87.72\% & 87.34\% & 87.34\% \\
        & log-mel-20 & 88.37\% & 87.86\% & \underline{91.46\%} & \underline{91.46\%} \\
        & log-mel-60 & 91.12\% & 91.12\% & \textbf{92.02\%} & \textbf{92.02\%} \\
        \bottomrule
    \end{tabular}
    }
    \caption{Audio-level performance comparison for pornographic sound detection. The results are obtained from the respective segment-level predictions using the various segment-to-audio prediction methods in \S\ref{ssec:segment-to-audio}.
    \vspace{-0.5cm}}
    \label{tab:result-per-audio}
\end{table*}

We assess the segment-level performance of each model and feature set pairs on both the validation and test set. We also measure their audio-level performances on the test set by employing segment-to-audio prediction methods (\S\ref{ssec:segment-to-audio}). We report the experimental results
in Table \ref{tab:result-per-segment} and Table \ref{tab:result-per-audio}. \textbf{Bold} denotes the best performance and \underline{underline} denotes the second best.

The CNN model with log mel spectrograms as its features and a 60-second segment length acquires the highest test F1-scores on the segment level (94.89\%) and audio level (92.02\%). Based on both experimental results, we conjecture that, in general: 1) the CNN manages to distinguish the pornographic sound features from the non-pornographic features and vice versa better than the FFNN; 2) log mel spectrograms are shown to be more discriminative than MFCCs in pornographic sound detection; and 3) appropriate segment length contributes to the model performance (using a 60-second segment consistently gives higher results than a 20-second segment). 

Moreover, we show the impact that the segment-to-audio prediction methods have on the model performance on the audio level. This is done by computing the average of all test F1-scores on the audio level by each method to compare the suitability of the segment-to-audio transformations, as shown in Table \ref{tab:result-per-audio}. The methods are then sorted from the highest to the lowest mean F1-score as follows:
\begin{enumerate*}[label=\arabic*)]
    \item voting (88.37\%),
    \item validation-dependent threshold (88.33\%),
    \item standard threshold (87.66\%),
    and
    \item segment majority (87.64\%)
\end{enumerate*}. As expected, the highest mean score is achieved by the voting method, which integrates the validation-dependent threshold and the segment majority. On its own, the performances derived from the validation-dependent threshold follow closely.

\subsection{Incorrect audio-level detection result analysis}
\label{ssec:incorrect-detection-analysis}

Out of 160 data samples in the test set classified by the best-performing model, only 13 are predicted incorrectly. To inspect further the possible causes of the erroneous classification, we manually examine the incorrectly predicted audio data. We find that false negatives can occur due to music or other continuous noise in the background in some parts of the audio data. In other cases, the audio data lack sex-related sounds, and mostly consist of environmental sounds (e.g., thunderstorm and water sounds) instead. These audio characteristics mislead the model to classify them as non-pornographic. The false positives. on the other hand, often consist of repeating sounds (e.g., clock ticking and hand claps) or human sounds (e.g., sighs and post-exercise heavy breathing). These sounds potentially cause the features derived from the non-pornographic audio to be deceptively similar to the pornographic audio.

\section{Conclusion and future work}
\label{sec:conclusion}


We present and compare various modeling methods for pornographic audio detection.
We conclude that CNNs trained on log mel spectrograms achieve the best performances on both the segment level (94.89\%) and audio level (92.02\%). We believe that this is due to the fact that log mel spectrograms are more fine-grained than MFCCs and the transforms performed by CNNs manage to extract the most important information from more complex feature representations~\cite{Loweimi2019}. Furthermore, we discover that the voting method for segment-to-audio prediction transformation is the most effective method to yield pornographic sound detection results on audio level.


 Since a DNN's performance relies heavily on the amount of training data, future research directions should concentrate on pornographic data augmentation. Another direction is to explore various acoustic features, such as the RCSFs used by \citet{lim2011automatic} and STDMFCCs used by \citet{kim2010automatic}, in conjunction with DNNs. Raw waveforms could also be used as lossless inputs to DNNs, allowing for even better discovery of sound artifacts. In the future, more complex DNNs, like the transformers, could be further explored as a pornographic sound modeling option.


\begin{acks}
This work is a part of "Intelligent System to Monitor Gadget Usage in Teenagers using Machine Learning Technique" research and funded by the Ministry of Research and Higher Education of Indonesia.
\end{acks}

\bibliographystyle{ACM-Reference-Format}
\bibliography{sample-base}

\end{document}